\documentclass[10pt,conference]{IEEEtran}

\newtheorem{proposition}{Proposition}
\newtheorem{corollary}{Corollary}
\newtheorem{definition}{Definition}
\newtheorem{lemma}{Lemma}
\usepackage{amsmath, amssymb,cite}
\usepackage[dvips]{graphicx}
\begin{document}
\newcommand{\df}{\stackrel{\mbox{\scriptsize def}}{=}}
\title{Properties of Codes with the Rank Metric}
\author{\authorblockN{Maximilien Gadouleau}
\authorblockA{Department of Electrical and Computer Engineering\\
Lehigh University\\
Bethlehem, PA 18015 USA} \and
\authorblockN{Zhiyuan Yan}
\authorblockA{Department of Electrical and Computer Engineering\\
Lehigh University\\
Bethlehem, PA 18015 USA}} \maketitle

\begin{abstract}
In this paper, we study properties of rank metric codes in general
and maximum rank distance (MRD) codes in particular. For codes with
the rank metric, we first establish Gilbert and sphere-packing
bounds, and then obtain the asymptotic forms of these two bounds and
the Singleton bound. Based on the asymptotic bounds, we observe that
asymptotically Gilbert-Varsharmov bound is exceeded by MRD codes and
sphere-packing bound cannot be attained. We also establish bounds on
the rank covering radius of maximal codes, and show that all MRD
codes are maximal codes and all the MRD codes known so far achieve
the maximum rank covering radius.
\end{abstract}
\IEEEpeerreviewmaketitle

\section{Introduction}
Early in the development of coding theory, it was found convenient
to model communication channels as conveyors of symbols from finite
sets and represent the effects of channel noise by occasional
reception of a symbol other than the transmitted symbol. Thus, the
Hamming metric has been considered the most relevant metric for
error-control codes. Recently, the rank metric has attracted some
attention due to its relevance to wireless communications
\cite{tarokh_98} and storage equipments \cite{roth_it91}. In
\cite{lusina_it03}, space-time block codes with good rank properties
have been proposed. Rank metric codes are used to correct crisscross
errors that can be found in memory chip arrays and magnetic tapes
\cite{roth_it91}. Codes with the rank metric have also been used in
the Gabidulin-Paramonov-Tretjakov (GPT) public-key cryptosystem
\cite{gabidulin_lncs91} and its variants (see, for example,
\cite{gabidulin_cc95,ourivski_dam03}). The public-key cryptosystems
based on codes with the rank metric have much smaller public key
sizes than those for Hamming metric based public-key cryptosystems
such as McEliece's cryptosystem \cite{mceliece_jet78}.

Due to these potential applications, general properties of codes
with the rank metric have received some attention
\cite{gabidulin_pit0185,roth_it91,babu_95,sripati_isit03,vasantha_gs99,vasantha_itw1002,vasantha_icads1202,manoj_report1002}.
The rank metric was considered as a metric for {\bf block} codes
over extension fields in \cite{gabidulin_pit0185}, whereas in
\cite{roth_it91} the rank metric was considered for {\bf array}
codes that consist of arrays over base fields. However, in both
\cite{gabidulin_pit0185} and \cite{roth_it91}, the same family of
codes that are optimal in the metric sense were proposed. In
\cite{sripati_isit03}, the rank distance properties of primitive
length linear cyclic codes were studied. In \cite{vasantha_gs99},
the rank covering radius of codes is studied, and the
sphere-covering bound for the rank metric is introduced. The concept
of rank covering radius is generalized in \cite{vasantha_itw1002},
where the multi-covering radii of codes with the rank metric are
defined. Recently, a somewhat more general construction for MRD
codes was proposed in \cite{kshevetskiy_isit05}, and the properties
of subspace subcodes and subfield subcodes were considered in
\cite{gabidulin_isit05}.

In this paper, we study properties of rank metric codes in general
and MRD codes in particular. For codes with the rank metric, we
first establish Gilbert and sphere-packing bounds, and then obtain
the asymptotic forms of these two bounds and the Singleton bound.
Based on these asymptotic bounds, we observe that MRD codes exceed
the Gilbert-Varsharmov bound, and that asymptotically perfect codes
(codes that attain the sphere-packing bound) do not exist. We also
establish bounds on the rank covering radius of maximal codes, and
show that all MRD codes are maximal codes and all MRD codes {\bf
known so far} achieve the maximum rank covering radius. The number
of vectors with certain weights determines the security against
decoding attacks of public-key cryptosystems based on error-control
codes. We compare the distributions of rank and Hamming weights of
vectors, and use the difference to partially explain why GPT
cryptosystem and its variants are quite secure against decoding
attacks.

The rest of the paper is organized as follows.
Section~\ref{sec:preliminaries} reviews necessary backgrounds in an
effort to make this paper self-contained. In
Section~\ref{sec:bounds}, we propose the Gilbert bound and the
sphere-packing bound for codes with the rank metric and their
asymptotic forms as well as the asymptotic form of the Singleton
bound. Section~\ref{sec:radius_gabidulin} establishes bounds on the
covering radius of maximal codes with the rank metric, and shows
that all the known MRD codes in the literature  achieve the maximum
rank covering radius. In Section~\ref{sec:weight}, the distributions
of rank and Hamming weights of vectors are compared and the
difference partially explains why GPT cryptosystem and its variants
are quite secure against decoding attacks.

\section{Preliminaries}\label{sec:preliminaries}
\subsection{Rank metric}
Consider $\mathbf{a} = (a_0, a_1,\ldots, a_{n-1}) \in
\mathrm{GF}(q^m)^n$, the $n$-dimensional vector space over
GF$(q^m)$. Assume $\gamma_0, \gamma_1, \ldots, \gamma_{m-1}$ is a
basis set of GF$(q^m)$ over GF$(q)$, then for $j=0, 1, \ldots, n-1$
$a_j$ can be written as $a_j = \sum_{i=0}^{m-1} a_{i, j}\gamma_i$,
where $a_{i, j} \in \mbox{GF}(q)$ for $i=0, 1, \ldots, m-1$. Hence,
$a_j$ can be expanded to an $m$-dimensional column vector $(a_{0,
j}, a_{1, j},\ldots, a_{m-1, j})^T$ with respect to the basis set
$\gamma_0, \gamma_1, \ldots, \gamma_{m-1}$. Let $\mathbf{A}$ be the
$m\times n$ matrix obtained by expanding all the coordinates of
$\mathbf{a}$. That is,
\begin{displaymath}
    \mathbf{A} = \left(
    \begin{array}{cccc}
        a_{0, 0} & a_{0, 1} & \ldots & a_{0, n-1}\\
        a_{1, 0} & a_{1, 1} & \ldots & a_{1, n-1}\\
        \vdots & \vdots & \ddots & \vdots\\
        a_{m-1, 0} & a_{m-1, 1} & \ldots & a_{m-1, n-1}
    \end{array}
    \right),
\end{displaymath}
where $a_j = \sum_{i=0}^{m-1} a_{i, j}\gamma_i$.  The \emph{rank
norm} (over GF$(q)$) of the vector ${\bf a}$, denoted as $rk({\bf
a}|\mbox{GF}(q))$, is defined to be the rank of the matrix ${\bf A}$
over GF$(q)$, i.e., $rk({\bf a}|\mbox{GF}(q)) \df rank({\bf A})$
\cite{gabidulin_pit0185}. Accordingly, $\forall \, {\bf a}, {\bf
b}\in \mbox{GF}(q^m)^n$, $d_{\mbox{\tiny R}}({\bf a},{\bf b})\df
rk({\bf a} - {\bf b}|\mbox{GF}(q))$ is shown to be a metric over
GF$(q^m)^n$, referred to as \emph{the rank metric} henceforth
\cite{gabidulin_pit0185}. Hence, the {\em minimum rank distance}
$d_{\mbox{\tiny R}}$ of a code of length $n$ is simply the minimum
rank distance over all possible pairs of distinct codewords.
Clearly, a code with a minimum rank distance $d_{\mbox{\tiny R}}$
can correct errors with rank up to $t = \left\lfloor (d_{\mbox{\tiny
R}}-1)/2 \right\rfloor$.

\subsection{The Singleton bound and MRD codes}
The minimum rank distance of a code can be specifically bounded.
First, the minimum rank distance $d_{\mbox{\tiny R}}$ of a code over
$\mathrm{GF}(q^m)$ is obviously bounded by $m$. Codes that satisfy
$d_{\mbox{\tiny R}} = m$ were referred to as full rank distance
codes and were studied in \cite{manoj_report1002}. Also, it can be
shown that $d_{\mbox{\tiny R}}\leq d_{\mbox{\tiny H}}$
\cite{gabidulin_pit0185}, where $d_{\mbox{\tiny H}}$ is the minimum
Hamming distance of the same code. Due to the Singleton bound for
block codes, the minimum rank distance of an $(n,k)$ block code over
$\mathrm{GF}(q^m)$ thus satisfies\footnote{For a nonlinear block
code over $\mathrm{GF}(q^m)$ with length $n$ and cardinality $M$,
the Singleton bound is similar: $d_{\mbox{\tiny R}}\leq
d_{\mbox{\tiny H}}\leq n-\left\lceil \log_{q^m} M \right\rceil +1$.}
\begin{equation}\label{eq:singleton1}
    d_{\mbox{\tiny R}}\leq n-k+1.
\end{equation}
An alternative bound on the minimum rank distance is also given in
\cite{loidreau_01}:
    \begin{equation}\label{eq:singleton2}
        d_{\mbox{\tiny R}} \leq \left\lfloor \frac{m}{n}(n-k) \right\rfloor + 1.
    \end{equation}
For $n \leq m$, the bound in (\ref{eq:singleton1}) is tighter than
that in (\ref{eq:singleton2}). When $n>m$ the bound in
(\ref{eq:singleton2}) is tighter. Since $\left\lfloor
\frac{m}{n}(n-k) \right\rfloor + 1 \leq m$ and the equality holds
only when $\frac{m\cdot k}{n}\leq 1$, the minimum rank distance of a
code must satisfy:
\begin{equation}\label{eq:singleton3}
    d_{\mbox{\tiny R}}\leq \min\left\{n-k+1, \left\lfloor \frac{m}{n}(n-k) \right\rfloor + 1\right\}.
\end{equation}
In this paper, we refer to the bound in (\ref{eq:singleton3}) as the
Singleton bound\footnote{The Singleton bound in \cite{roth_it91} has
a different form since array codes are defined over base fields.}
for codes with the rank metric, and call codes that attain the bound
as maximum rank distance (MRD) codes. The Singleton bound for codes
with the rank metric implies the rate-diversity tradeoff in
\cite{LK03}.

Three subclasses of MRD codes have been proposed to our best
knowledge. The first subclass of MRD codes, called Gabidulin codes,
was first introduced in \cite{gabidulin_pit0185}.
\begin{definition}[Gabidulin codes]\label{def:gabidulin}
When $n \leq m$, let $\mathbf{g} = (g_0, g_1, \ldots, g_{n-1})$ be
linearly independent elements of $\mathrm{GF}(q^m)$. Then the code
defined by the following generator matrix
\begin{equation}
    \mathbf{G} = \left( \begin{array}{cccc}
    g_0 & g_1 & \ldots & g_{n-1}\\
    g_0^{[1]} & g_1^{[1]} & \ldots & g_{n-1}^{[1]}\\
    \vdots & \vdots & \ddots & \vdots\\
    g_0^{[k-1]} & g_1^{[k-1]} & \ldots & g_{n-1}^{[k-1]}
    \end{array}
    \right),\label{eq:gabidulin}
\end{equation}
where $[i] = q^i$, is called a \emph{Gabidulin code}, generated by
$\mathbf{g} = (g_0, g_1, \ldots, g_{n-1})$, with dimension $k$ and
minimum rank distance $d_{\mbox{\tiny R}} = n-k+1$.
\end{definition}
A second subclass of MRD codes, referred to as generalized Gabidulin
codes, was recently introduced in \cite{kshevetskiy_isit05}. These
codes have a similar generator matrix to that in
(\ref{eq:gabidulin}) except that for this subclass of codes $[i] =
q^{ai}$ with $a$ being an integer prime to $m$. Even though
Gabidulin codes are only a special case of generalized Gabidulin
codes (for $a=1$), we consider Gabidulin codes as a separate
subclass since their properties have been studied more extensively.
The third subclass of MRD codes consists of cartesian products of an
MRD code with length $n = m$. Let $C$ be an $(n, k, d_{\mbox{\tiny
R}}=n-k+1)$ MRD code over GF$(q^m)$ ($n \leq m$), and let $C^l \df C
\times \ldots \times C$ be the code obtained by $l$ cartesian
products of $C$. Thus, $C^l$ is an code with length $n' = nl$,
dimension $k' = kl$, and minimum rank distance $d_{\mbox{\tiny R}}'
= d_{\mbox{\tiny R}}=n-k+1$. It can be shown that $C^l$ is an MRD
code if and only if $n=m$. The first two subclasses of MRD codes
have length less than $m$, whereas this third subclass consists of
codes with length $n'=lm \geq m$.

\section{Bounds for the rank metric}\label{sec:bounds}
\subsection{Gilbert and sphere-packing bounds for the rank metric}
Let us denote the number of $n$-dimensional vectors with rank weight
$w$ over GF$(q^m)$ as $N_{q^m}(n,w)$, it is given by
\begin{equation} N_{q^m}(n,w) =\prod_{i=0}^{w-1}\frac{(q^n-q^i)(q^m-q^i)}{(q^w-q^i)}.
\label{eq:vectors}
\end{equation} Denoting the volume of a ball of
radius $w$ (in the rank metric) in GF$(q^m)^n$ by $V_{q^m}(n,w)$, we
obtain
\begin{equation} V_{q^m}(n,w) = \sum_{j=0}^w
N_{q^m}(n,j).\label{eq:volume}\end{equation}

Bounds on the parameters of codes indicate how good codes are, and
also provide guidelines to the design of good codes. The Gilbert and
sphere-packing bounds are two important bounds for codes in the
Hamming metric \cite{blahut_83,macwilliams_77}. First, the Gilbert
bound states that there always exists a code with $A$ codewords,
length $n$, minimum distance $d$ such that $A \geq
\frac{q^{mn}}{V_{q^m}(n,d-1)}$. The sphere-packing bound, on the
other hand, states that any code has to satisfy $A \leq
\frac{q^{mn}}{V_{q^m}(n,t)}$, where $t = \left\lfloor (d-1)/2
\right\rfloor$. The derivations of these two bounds do not depend on
the metric considered, and hence these two bounds can be easily
adapted to codes with the rank metric. Let us denote the maximum
number of codewords in a code of length $n$ and minimum rank
distance $d_{\mbox{\tiny R}}$ over GF$(q^m)$ as $A_{q^m}(n,
d_{\mbox{\tiny R}})$. The Gilbert and sphere-packing bounds for
codes with the rank metric are given by
\begin{equation}
    \frac{q^{mn}}{V_{q^m}(n,d_{\mbox{\tiny R}}-1)} \leq A_{q^m}(n,
    d_{\mbox{\tiny R}}) \leq
    \frac{q^{mn}}{V_{q^m}(n,t)}.\label{ieq:bounds}
\end{equation}

The Gilbert bound gives a lower bound to the cardinality of a
``reasonably good'' code for given block length and minimum
distance. More formally, one can show that the Gilbert bound is
always reached or exceeded by a special class of codes, called
maximal codes.
\begin{definition}[Maximal code]
A code $C$ with length $n$ and minimum rank distance $d_{\mbox{\tiny
R}}$ is maximal if there does not exist any code $C'$ with same
length and minimum rank distance such that $C \subset C'$.
\end{definition}

We can show that
\begin{proposition}\label{prop:maximal_gabidulin}
All MRD codes are maximal codes.
\end{proposition}
Due to limited space, all proofs are omitted in this paper and they
will be presented at the conference.
Proposition~\ref{prop:maximal_gabidulin} implies that Gabidulin
codes, generalized Gabidulin codes, and cartesian products of MRD
codes of length $m$ are all maximal codes. However, it is not the
case for all cartesian products of Gabidulin (or generalized
Gabidulin) codes. Indeed, we can show that
\begin{proposition}\label{prop:maximal_cartesian}
Let $C$ be an $(n, k, d_{\mbox {\tiny R}})$ MRD code over GF$(q^m)$
($n \leq m$). If $l \geq \frac{m}{m-n}$ and $d_{\mbox {\tiny R}} >
1$, then $C^l$ is not a maximal code.
\end{proposition}

\subsection{Asymptotic bounds for the rank metric}
The performance of codes of large block length can be studied in
terms of asymptotic bounds on the relative minimum distance in the
limit of infinite block length. In this section, we will study the
asymptotic forms for the three bounds in (\ref{eq:singleton3}) and
(\ref{ieq:bounds}) respectively in the case where both block length
and minimum rank distance go to infinity. However, this cannot be
achieved for finite $m$ since the minimum rank distance is no
greater than $m$. Thus, we consider the case where $\lim_{n
\rightarrow \infty}\frac{n}{m}=b$, where $b$ is a constant.

Define
$$
    \delta \df \frac{d_{\mbox{\tiny R}}}{n} \mbox{ and } a(\delta) \df \lim_{n\rightarrow
    \infty}\sup \left[ \frac{\log_{q^m}A_{q^m}(n,\lfloor\delta
    n\rfloor)}{n}\right],
$$
where $a(\delta)$ represents the maximum possible code rate of a
code which has relative minimum distance $\delta$ as its length goes
to infinity. The asymptotic forms of the bounds
in~(\ref{ieq:bounds}) are given in the propositions below.

\begin{proposition}[Gilbert-Varsharmov bound]
\label{prop:gv_bound}
The asymptotic behavior of the Gilbert bound
for the rank metric is given by (for $0 \leq \delta \leq
\min\{1,b^{-1}\}$)
\begin{equation} a(\delta) \geq (1-\delta)(1-b\delta),\label{ieq:gv}
\end{equation}
which will be referred to as the Gilbert-Varsharmov bound for the
rank metric.
\end{proposition}

\begin{proposition}[Asymptotic sphere-packing bound]
The asymptotic behavior of the sphere-packing bound for the rank
metric is given by (for $0 \leq \delta \leq \min\{1,b^{-1}\}$)
\begin{equation}a(\delta) \leq
\left(1-\frac{\delta}{2}\right)\left(1-b\frac{\delta}{2}\right).
\label{ieq:sphere}
\end{equation}
\end{proposition}

The rank of an $n \times bn$ matrix is equal to the rank of its
transpose. Hence, for a code $C$ of length $n$ and relative minimum
distance $\delta$ over GF$(q^{bn})$, the transpose code $C^{T}$ of
length $bn$ over GF$(q^{n})$ has relative minimum distance $\delta'
= b\delta$. Therefore, changing $b$ into $b^{-1}$ and $\delta$ into
$b\delta$ will leave the bounds in (\ref{ieq:gv}) and
(\ref{ieq:sphere}) unchanged.

The Singleton bound for the rank metric in (\ref{eq:singleton3})
asymptotically becomes
\begin{equation}
    a(\delta) \leq \left\{\begin{array}{ll}
    1-\delta & \mbox{if}\,b\leq 1,\\
    1-b\delta & \mbox{if}\,b > 1.
    \end{array}\right.\label{ieq:singleton}
\end{equation}

\begin{figure}[htp]
\begin{center}
\includegraphics[scale=0.62]{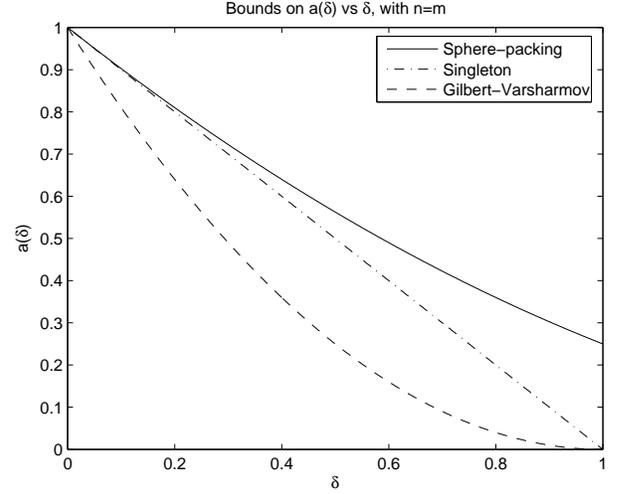}
\end{center}
\caption{The sphere-packing, Singleton, and Gilbert-Varsharmov
asymptotic bounds for the rank metric, with
$n=m$.}\label{fig:bounds1}
\end{figure}

\begin{figure}[htp]
\begin{center}
\includegraphics[scale=0.62]{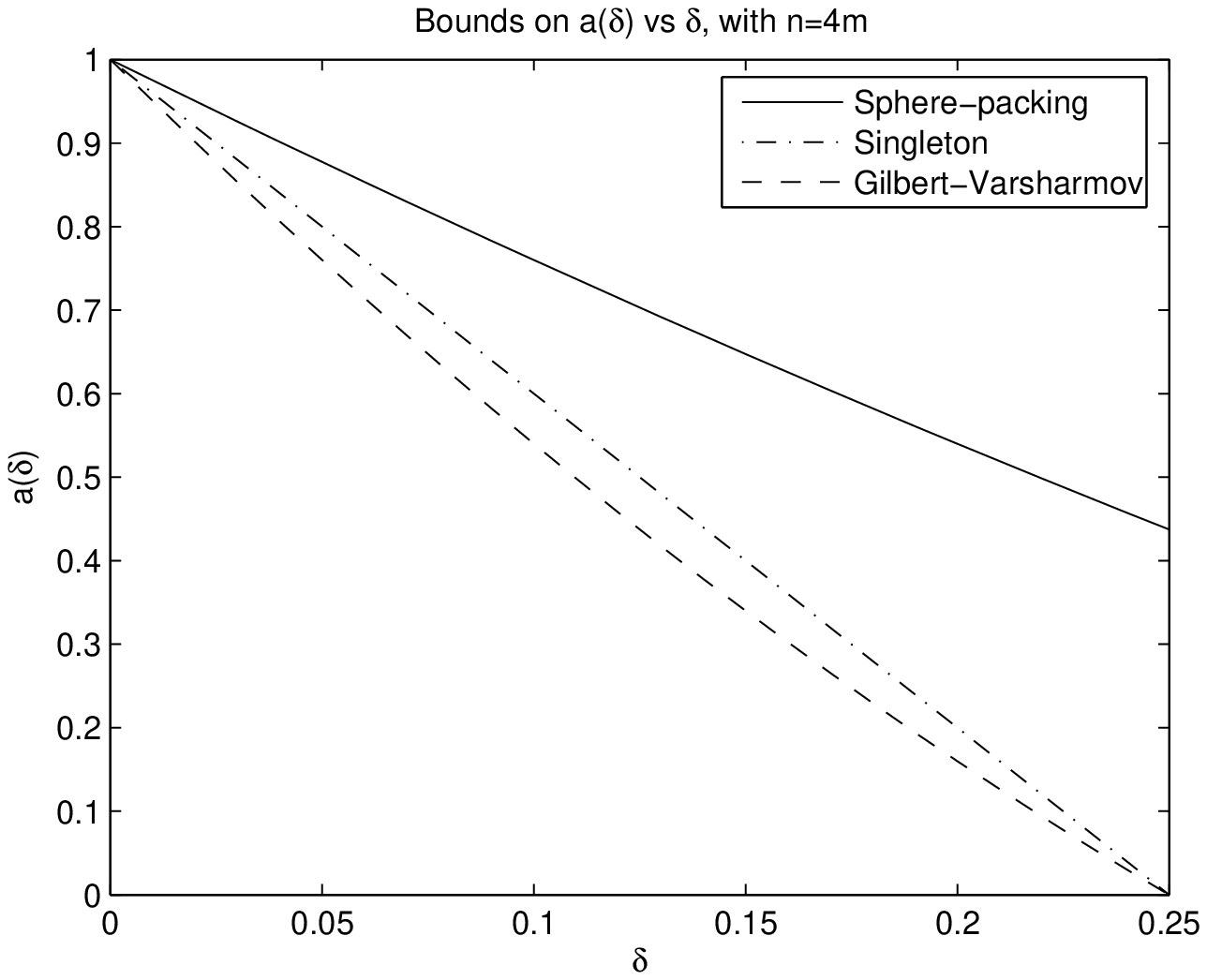}
\end{center}
\caption{The sphere-packing, Singleton, and Gilbert-Varsharmov
asymptotic bounds for the rank metric, with
$n=4m$.}\label{fig:bounds4}
\end{figure}

\begin{figure}[htp]
\begin{center}
\includegraphics[scale=0.62]{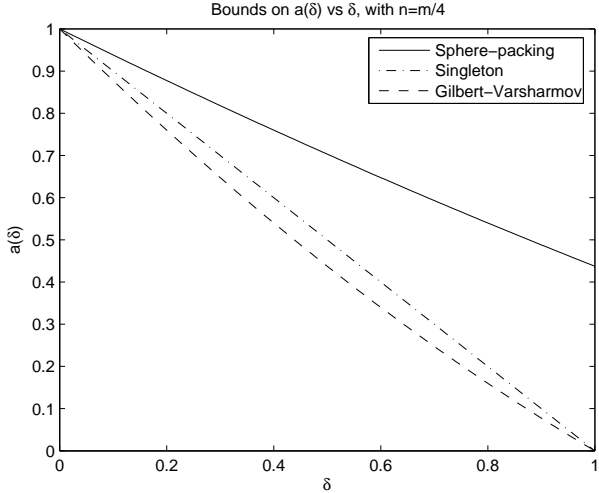}
\end{center}
\caption{The sphere-packing, Singleton, and Gilbert-Varsharmov
asymptotic bounds for the rank metric, with
$n=\frac{1}{4}m$.}\label{fig:bounds0.25}
\end{figure}

The three asymptotic bounds are illustrated in
Figures~\ref{fig:bounds1}, \ref{fig:bounds4}, and
\ref{fig:bounds0.25} for $b=1, 4, \mbox{ and } 0.25$, respectively.
Insights about asymptotic behavior of codes in the rank metric can
be obtained from these asymptotic bounds. First, note that for
$\delta > 0$ the sphere-packing bound is always looser than the
Singleton bound. In particular, when $b=1$ and $\delta =1$, then the
right hand side of (\ref{ieq:sphere}) is $1/4$, although $a(\delta)
= 0$. Since both the sphere-packing and Singleton bounds are upper
bounds, this implies that the sphere-packing bound cannot be
attained asymptotically. That is, asymptotically there are no
perfect codes in the rank metric. This confirms the claim in
\cite{babu_95} that there are no perfect codes for the rank metric.

The values $0, 1, \mbox{ and } \infty$ are the special cases for
$b$. When $b\rightarrow 0$, the right hand sides of (\ref{ieq:gv})
and (\ref{ieq:singleton}) coincide. This means that when $m$
increases faster than linearly with $n$, then MRD codes do not
exceed the Gilbert-Varsharmov. This is similar to the case where the
Gilbert-Varsharmov bound for codes with the Hamming metric is
reached but not exceeded by MDS codes. When $b=1$, it can be shown
that the gap between the Singleton bound and Gilbert-Varsharmov
bound is maximized. When $b\rightarrow  \infty$, the bounds in
(\ref{ieq:gv}), (\ref{ieq:sphere}), and (\ref{ieq:singleton}) are
valid only at the point $\delta=0$. This is because $\lim_{n
\rightarrow \infty}\frac{d_{\mbox{\tiny R}}}{n}\leq \lim_{n
\rightarrow \infty}\frac{m}{n}=0$. The presence of critical points
for $b \rightarrow 0$ and $b \rightarrow \infty$ confirms our choice
of studying the case where the ratio $\frac{n}{m}$ tends to a real
numbered value.

Since there exist MRD codes for any value of $\delta$, the bound
in~(\ref{ieq:singleton}) is attained by MRD codes. We note that the
asymptotic code rates of MRD codes are always greater than
$(1-\delta)(1-b\delta)$ unless $b\rightarrow 0$. Therefore, MRD
codes exceed the Gilbert-Varsharmov bound for the rank metric
unless $b\rightarrow 0$, as illustrated in
Figures~\ref{fig:bounds1},~\ref{fig:bounds4}
and~\ref{fig:bounds0.25}.

Let us also study the asymptotic behavior of cartesian products of
MRD codes. Let $C$ be an $(n, k, d_{\mbox{\tiny R}})$ MRD code over
GF$(q^m)$ ($n \leq m$), and let $C^l \df C \times \ldots \times C$
be the code obtained by $l$ cartesian products of $C$ with length
$n' = nl$, dimension $k' = kl$, and minimum rank distance
$d_{\mbox{\tiny R}}' = d_{\mbox{\tiny R}} = n-k+1$. Clearly, we have
$k' = n' - l(d_{\mbox{\tiny R}}'-1)$. Thus, $\lim_{n \rightarrow
\infty}\frac{k'}{n'} = 1 - l\delta$ for $0 \leq \delta \leq l^{-1}$.
Let us define $b= \lim_{n \rightarrow \infty}\frac{n'}{m}$, then we
have $l^{-1} \leq b^{-1}$. Hence, $C^l$ reaches or exceeds the
Gilbert-Varsharmov bound if and only if $\delta \leq
\frac{b+1-l}{b}$. We can show that when $l \geq b+1$, $C^l$ does not
attain the Gilbert-Varsharmov bound and hence is not asymptotically
maximal for any $\delta > 0$. This confirms the result in
Proposition~\ref{prop:maximal_cartesian}.

\section{Covering radius}\label{sec:radius_gabidulin}

Let $C$ be an $(n,k)$ code over GF$(q^m)$. The covering radius in
the rank metric $r(C)$ of this code is defined in
\cite{vasantha_gs99} similarly to the covering radius in the Hamming
metric. It is the smallest integer $r$ such that all vectors in the
space GF$(q^m)^n$ are within rank distance $r$ of some codeword. The
covering radius is an important geometric property of a code: it is
a measure of the maximum distortion if the code is used for data
compression, and is the maximum weight of a correctable random error
if the code is used for error correction \cite{cohen_it0585}.

\subsection{General properties of the covering radius}
From the definition of the covering radius, it is clear that for any
code $C$, $r(C) \leq m$. Also, if we further assume that $C$ is
linear, $r(C)$ is bounded by $n-k$ \cite{vasantha_gs99}. Similarly
to the Hamming covering radius, the rank covering radius of a
maximal code $C$ satisfies $r(C) \leq d(C) - 1$. Combining this
result with the Singleton bound for the rank metric
in~(\ref{eq:singleton3}), we obtain that for any (linear or
nonlinear) maximal code $C$,
\begin{equation}\label{eq:radius}
    r(C) \leq \min \left\{ n-k, \left\lfloor\frac{m}{n}(n-k)\right\rfloor\right\}.
\end{equation}
Note that the bound in (\ref{eq:radius}) is not applicable to
general codes with the rank metric. A trivial counter example is
given by the $(n,1)$ repetition code with length $n \geq m+1$, which
has covering radius $n-1$ \cite{vasantha_gs99} that exceeds the
bound in (\ref{eq:radius}).

\subsection{Covering radius of MRD codes}
If we assume that $n \leq m$, then (\ref{eq:radius}) becomes $r(C)
\leq n-k$. In the following, we show that generalized Gabidulin
codes have the maximum covering radius of $n-k$. The proof follows
the same arguments used in the derivation of the Hamming covering
radius of Reed-Solomon codes \cite{cohen_it0585}.

\begin{definition}
Let $C_1 \subset C_2$ be two codes. We denote by $m(C_2,C_1)$ [and
$M(C_2, C_1)$] the weight of the translate-leader of least
[greatest] nonzero rank weight among the translates of $C_1$ by
elements of $C_2$, i.e.,
\begin{eqnarray}
    m(C_2,C_1) & = & \min_{{\bf x} \in C_2 - C_1}\{w({\bf x} + {\bf c}) \,|\, {\bf c} \in
    C_1 \}, \\
    M(C_2,C_1) & = & \max_{{\bf x} \in C_2} \min\{w({\bf x} + {\bf c}) \,|\, {\bf c} \in
    C_1\}.
\end{eqnarray}
When $C_1$ and $C_2$ are linear, these are the minimum nonzero and
maximum weights of cosets of $C_2\, \mbox{mod}\,C_1$.
\end{definition}
We remark that $M(C_2, C_1) \geq m(C_2,C_1)$, and that if $C_2
\subseteq C_3$, then $M(C_3,C_1) \geq M(C_2,C_1)$.
\begin{lemma}[Supercode Lemma]\label{lemma:supercode}
Let $C_1$ and $C_2$ be two linear codes such that $C_1 \subset C_2$.
Then $r(C_1) \geq M(C_2, C_1) \geq m(C_2,C_1) \geq d(C_2)$. Also,
$$
r(C_1) \geq \min_{{\bf x} \in C_2-C_1}\{w({\bf x})\}.
$$
\end{lemma}
Using Lemma~\ref{lemma:supercode}, we can show that generalized
Gabidulin codes have maximal covering radius.
\begin{proposition}\label{prop:radius_gabidulin}
An $(n,k,d_{\mbox{\tiny R}})$ generalized Gabidulin code over
GF$(q^m)$ ($m \geq n$) has covering radius $d_{\mbox{\tiny
R}}-1=n-k$.
\end{proposition}

A similar argument can be used to bound the covering radius of the
cartesian products of generalized Gabidulin codes.
\begin{corollary}\label{prop:radius_cartesian}
Let $C$ be an $(n,k,d_{\mbox{\tiny R}})$ generalized Gabidulin code
$(n \leq m)$, and let $C^l$ be the $(n',k',d_{\mbox{\tiny R}}')$
code obtained by $l$ cartesian products of $C$. Then, $\forall\, l
\geq 1$, the rank covering radius of $C^l$ satisfies $r(C^l) \geq
d_{\mbox{\tiny R}}'-1$.
\end{corollary}
Note that we do not have an equality as in
Corollary~\ref{prop:radius_gabidulin} since cartesian products of
MRD codes are not necessarily maximal, as stated in
Proposition~\ref{prop:maximal_cartesian}. However, when $n=m$, $C^l$
is an MRD code, therefore its covering radius also satisfies $r(C^l)
\leq d(G^l) - 1$. This leads to the following result.
\begin{corollary}
Let $C$ be an $(m,k,d_{\mbox{\tiny R}})$ generalized Gabidulin code
over GF$(q^m)$, and let $C^l$ be the $(n',k',d_{\mbox{\tiny R}}')$
code obtained by $l$ cartesian products of $C$. Then $r(C^l) =
d_{\mbox{\tiny R}}'-1 = \frac{m}{n'}(n'-k')$.
\end{corollary}

In summary, we conclude that all three subclasses of MRD codes known
so far have maximal covering radius. Note that for the Hamming
metric however, it is known that some MDS codes do not have maximal
covering radius \cite{cohen_it0585}.

\section{Rank weight distribution}\label{sec:weight}
\begin{figure}[htb]
\begin{center}
\includegraphics*[scale=0.6]{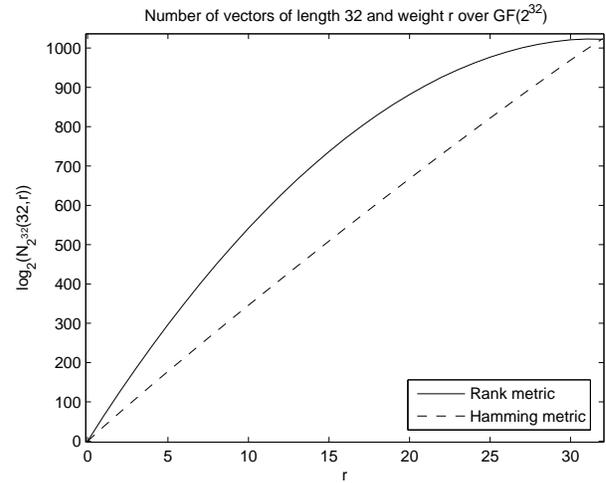}
\caption{The number of vectors with length $32$ and weight $r$ over
GF$(2^{32})$\label{fig:vectors}}
\end{center}
\end{figure}

All public-key cryptosystems based on error codes encrypt the
plaintext by first encoding it using the public code and then adding
an error vector of weight $t$ (see, for example,
\cite{mceliece_jet78,gabidulin_lncs91} for details). Thus, a brute
force decoding attack \cite{mceliece_jet78} can be used to break
these cryptosystems: first guess the error vector of weight $t$, and
then subtract the guessed error vector and invert the encoding
process; the system is broken if the guessed error vector is the
error vector used in the encryption operation, otherwise repeat with
a different guess. Clearly, the work factor of such a decoding
attack is proportional to the number of the vectors with weight $t$,
$N_{q^m}(n, t)$. For codes with the rank metric, it is given in
(\ref{eq:vectors}). For codes with the Hamming metric, it is given
by $(q^m-1)^t \binom{n}{t}$. It can be shown that for $0<r<n$ the
number of the vectors with the rank weight $r$ is much greater than
the number of the vectors with the Hamming weight $r$. The numbers
of vectors with length $32$ and rank and Hamming weight $r$ ($0\leq
r \leq 32$), respectively, over GF$(2^{32})$ are compared in
Figure~\ref{fig:vectors}. Thus, the public-key cryptosystems based
on rank metric codes are more secure against this brute force
decoding attack. This partially explains why the GPT cryptosystem
and its variants are quite secure against decoding attacks.

\bibliographystyle{IEEEtran}
\bibliography{gpt}

\end{document}